\documentclass[conference]{IEEEtran}
\usepackage[edges]{forest}
\usetikzlibrary{shadows}
\IEEEoverridecommandlockouts

\usepackage[normalem]{ulem}
\usepackage{listings}
\usepackage{float}
\usepackage{xcolor}
\usepackage{array}
\usepackage{balance}
\usepackage{enumitem}
\usepackage[T1]{fontenc}
\usepackage{beramono, tikz}
\usepackage[flushleft]{threeparttable}

\definecolor{folderbg}{RGB}{124,166,198}
\definecolor{folderborder}{RGB}{110,144,169}

\def\Size{4pt}
\tikzset{
  folder/.pic={
    \filldraw[draw=folderborder,top color=folderbg!50,bottom color=folderbg]
      (-1.05*\Size,0.2\Size+5pt) rectangle ++(.75*\Size,-0.2\Size-5pt);  
    \filldraw[draw=folderborder,top color=folderbg!50,bottom color=folderbg]
      (-1.15*\Size,-\Size) rectangle (1.15*\Size,\Size);  }
}

\definecolor{codegreen}{rgb}{0,0.6,0}
\definecolor{codegray}{rgb}{0.5,0.5,0.5}
\definecolor{codepurple}{rgb}{0.58,0,0.82}
\definecolor{backcolour}{rgb}{0.95,0.95,0.92}

\makeatletter
\newenvironment{btHighlight}[1][]
{\begingroup\tikzset{bt@Highlight@par/.style={#1}}\begin{lrbox}{\@tempboxa}}
{\end{lrbox}\bt@HL@box[bt@Highlight@par]{\@tempboxa}\endgroup}

\newcommand\btHL[1][]{%
  \begin{btHighlight}[#1]\bgroup\aftergroup\bt@HL@endenv%
}
\def\bt@HL@endenv{%
  \end{btHighlight}%
  \egroup
}
\newcommand{\bt@HL@box}[2][]{%
  \tikz[#1]{%
    \pgfpathrectangle{\pgfpoint{1pt}{0pt}}{\pgfpoint{\wd #2}{\ht #2}}%
    \pgfusepath{use as bounding box}%
    \node[anchor=base west, fill=orange!30,outer sep=0pt,inner xsep=1pt, inner ysep=0pt, rounded corners=3pt, minimum height=\ht\strutbox+1pt,#1]{\raisebox{1pt}{\strut}\strut\usebox{#2}};
  }%
}
\makeatother

\lstdefinestyle{Cpp}{
    commentstyle=\scriptsize\color{codegreen},
    keywordstyle=\scriptsize\color{codegreen},
    numberstyle=\scriptsize\color{codegray},
    stringstyle=\scriptsize\color{codepurple},
    extendedchars=true,
    basicstyle=\scriptsize\ttfamily,
    breakatwhitespace=false,         
    breaklines=true,                 
    captionpos=b,                    
    keepspaces=true,                 
    numbers=left,                    
    numbersep=5pt,                  
    showspaces=false,                
    showstringspaces=false,
    showtabs=false,                  
    tabsize=2,
    moredelim=**[is][\btHL]{@}{@}
}

\usepackage{cite}
\usepackage{soul}
\usepackage{amsmath,amssymb,amsfonts}
\usepackage{mathtools}
\usepackage[noend]{algorithmic}
\usepackage[ruled, vlined, linesnumbered]{algorithm2e}
\usepackage{graphicx, changepage}
\usepackage{svg}
\usepackage{caption}
\usepackage{textcomp}
\usepackage{hyperref}
\usepackage{booktabs}
\usepackage{authblk}
\usepackage{vcell}
\usepackage{multirow}
\usepackage{balance}
\def\BibTeX{{\rm B\kern-.05em{\sc i\kern-.025em b}\kern-.08em
    T\kern-.1667em\lower.7ex\hbox{E}\kern-.125emX}}

\makeatletter 
\newcommand{\linebreakand}{%
  \end{@IEEEauthorhalign}
  \par
  \hfill\begin{@IEEEauthorhalign}
}
\makeatother 

\begin{document}

\title{Are We Learning the Right Features?\\A Framework for Evaluating DL-Based\\Software Vulnerability Detection Solutions
}

\author[]{Satyaki Das}
\author[]{Syeda Tasnim Fabiha}
\author[]{Saad Shafiq}
\author[]{Nenad Medvidović}
\affil[]{University of Southern California
\authorcr {\{satyakid, fabiha, sshafiq, neno\}@usc.edu}}




\maketitle

\begin{abstract}
\looseness-1
Recent research has revealed that the reported results of an emerging body of deep learning-based techniques for detecting software vulnerabilities are not reproducible, either across different datasets or on unseen samples. This paper aims to provide the foundation for properly evaluating the research in this domain. We do so by analyzing prior work and existing vulnerability datasets for the syntactic and semantic features of code that contribute to vulnerability, as well as features that falsely correlate with vulnerability. We provide a novel, uniform representation to capture both sets of features, and use this representation to detect the presence of both vulnerability and spurious features in code. To this end, we design two types of code perturbations: feature preserving perturbations (FPP) ensure that the vulnerability feature remains in a given code sample, while feature eliminating perturbations (FEP) eliminate the feature from the code sample. These perturbations aim to measure the influence of spurious and vulnerability features on the predictions of a given vulnerability detection solution. To evaluate how the two classes of perturbations influence predictions, we conducted a large-scale empirical study on five state-of-the-art DL-based vulnerability detectors. Our study shows that, for vulnerability features, only \textasciitilde2\% of FPPs yield the undesirable effect of a prediction changing among the five detectors on average. However, on average, \textasciitilde84\% of FEPs yield the undesirable effect of retaining the vulnerability predictions. For spurious features, we observed that FPPs yielded a drop in recall up to 29\% for graph-based detectors. We present the reasons underlying these results and suggest strategies for improving DNN-based vulnerability detectors. We provide our perturbation-based evaluation framework as a public resource to enable independent future evaluation of vulnerability detectors.
\end{abstract}

\begin{IEEEkeywords}
vulnerability detection, deep learning, software security, explainable AI
\end{IEEEkeywords}

\section{Introduction}
\label{sec:introduction}

Identifying security vulnerabilities in software, and specifically in source code, has been an important focus of researchers and practitioners, prompted by numerous examples of high-profile security breaches~\cite{li2018vuldeepecker,bhatt2017modeling, grobauer2010understanding, piessens2002taxonomy, santos2017understanding, sejfia2024toward}. Earlier research in this area concentrated on developing deterministic approaches for vulnerability detection that relied on predefined rules and patterns~\cite{pham2010detection, shin2010evaluating, li2016vulpecker, kim2017vuddy}. Since these approaches have suffered from a range of shortcomings~\cite{li2018vuldeepecker}, researchers have more recently turned to deep learning as a vehicle for vulnerability detection because DL offers a superior capacity to learn complex patterns from data~\cite{li2018vuldeepecker,li2021sysevr,cheng2021deepwukong,chakraborty2021deep, fu2022linevul}. DL techniques have demonstrated their versatility in other software engineering tasks that involve source code datasets, such as code clone detection and authorship attribution~\cite{cito2022counterfactual}, providing additional motivation for their use in software vulnerability detection.

\looseness=-1
Although the shift to DL has yielded promising results, it has also introduced new challenges. Specifically, these techniques operate as black boxes, making it difficult to understand the reasoning behind their predictions and decisions. They also suffer from a lack of generalizability~\cite{ribeiro2016should,chakraborty2021deep}, performing poorly on unseen datasets and failing to adapt to new vulnerabilities. It is thus important to make these techniques more explainable, and this can be achieved by investigating the specific code features that influence their predictions. By doing so, we can uncover the underlying decision-making processes, expose potential biases and limitations, and pinpoint areas for refinement, ultimately leading to the development of more trustworthy, reliable, and effective vulnerability detection techniques. 

On this front, existing literature has recognized the presence of spurious features in DL-based approaches~\cite{rahman2023towards,risse2023limits,chakraborty2021deep}. These are code features that falsely correlate with the target label. Such spurious features can impact vulnerability detection tools and models, and they provide a helpful starting point for our work. However, to systematically advance the state-of-the-art in vulnerability detection, a three-pronged approach is necessary: (1)~identify and disregard spurious features~(\textit{SF}) in code that can lead to inaccuracies; (2)~pinpoint and leverage genuine  features that contribute to vulnerabilities~(\textit{VF}); and (3)~analyze and quantify the impact of \textit{SF} and \textit{VF} on a proposed vulnerability detection technique. 

\looseness=-1
This paper presents our implementation of the above three-pronged approach. First, we conducted a rigorous analysis of the widely used SARD vulnerability dataset~\cite{sard} to uncover the key features and patterns that contribute to the manifestation of the vulnerabilities in the dataset~\textit{VF}. 
Second, we expanded the list of \textit{SF}s by exploring the assumptions made in the literature (e.g., those that do not hold true in our dataset samples).
We have systematized the uncovered \textit{VF} and \textit{SF} and structured them into an expandable taxonomy of code features for vulnerability detection. 

Third, we have developed \textit{VIPer}, a novel perturbation-based approach for identifying the weaknesses in a given vulnerability detector's predictions. 
\textit{VIPer} generates both feature preserving perturbations (FPP), which ensure that a feature (\textit{VF} or \textit{SF}) remains in a given code sample, and feature eliminating perturbations (FEP), which remove the feature from the code sample. \textit{VIPer} comprises three phases: (1)~\textit{Feature detection} identifies the presence or absence of each feature from our taxonomy in a given source code sample. 
(2)~\textit{Targeted perturbation} modifies the code sample in a manner that either preserves (FPP) or removes (FEP) a detected feature. 
(3)~\textit{Solution evaluation} involve analyzing a vulnerability detector's response to the targeted perturbations and inferring the extent to which a given \textit{VF} or \textit{SF} impacts the detector's prediction. 

We have applied \textit{VIPer} on five state-of-the-art DL-based vulnerability detectors: DeepWukong~\cite{cheng2021deepwukong}, ReVeal~\cite{chakraborty2021deep}, DeepDFA~\cite{steenhoek2024dataflow}, LineVul~\cite{fu2022linevul}, and SySeVR~\cite{li2021sysevr}. By analyzing the five detectors' responses to \textit{VIPer}'s perturbations, we quantified the extent to which a given feature contributes to a prediction, thus providing valuable insights into the detectors' decision-making processes, the sensitivity and robustness of their predictions, and the potential biases and limitations of their vulnerability detection capabilities. 
Our findings indicate that, in case of \textit{VF}s \textit{VIPer}'s perturbations significantly impact the five detectors' performance, with precision decreasing by \textasciitilde28\% and recall  by \textasciitilde8\%, on average. The detectors exhibit reasonable robustness to FPPs, with only \textasciitilde2\% of all FPPs yielding the inappropriate outcome of changed vulnerability predictions. However, \textasciitilde84\% of FEPs result in the inappropriate outcome of retained vulnerability predictions. Additionally, in the case of \textit{SF}s, FPPs produce a decline in recall of up to 29\%. Together, the latter two results mean that, in an overwhelming majority of cases, the state-of-the-art vulnerability detectors' original reasoning behind predictions was flawed as it was not actually based on the targeted features.

This paper makes the following contributions:
\begin{itemize}
    \item an extendable taxonomy of vulnerability (\textit{VF}) and spurious (\textit{SF}) features; 
    \item \textit{VIPer}, a perturbation-based framework~\cite{zenodo} to gauge the robustness of vulnerability detectors;
    
    
    \item a customizable wrapper for seamless integration of the framework in the evaluation pipeline of existing vulnerability detectors; and
    \item a comprehensive empirical evaluation of five state-of-the-art vulnerability detectors, assessing the impact of both \textit{VF} and \textit{SF} on their predictions.
\end{itemize}

In the paper's remainder, Section~\ref{sec:definitions} introduces the novel taxonomy of code features. Section~\ref{sec:approach} details our approach, \textit{VIPer}. Sections~\ref{sec:exp} and \ref{sec:results} present the evaluation setup and results of our study. Section~\ref{sec:discuss} discusses our findings and their implications. Threats to validity are discussed in Section~\ref{sec:threats}, related work in Section~\ref{sec:lit}, and conclusions in Section~\ref{sec:conclusion}.

\section{Taxonomy of Code Features}
\label{sec:definitions}
{
We initiated our study by examining which code features a given detector learns, by analyzing the widely-adopted vulnerability datasets: SARD~\cite{sard}, FFmpeg+Qemu~\cite{zhou2019devign}, Draper~\cite{russell2018automated}, and BigVul~\cite{fan2020ac}. All four datasets were instructive in our understanding of the problem and its different manifestations. However, only SARD provided annotations at critical points in the source code that describe how a vulnerability manifests itself in the vulnerable sample (e.g., see comment prefixed with ``FLAW'' on line 3 in Listing~\ref{lst:vul_smp}) and what changes one can apply to repair it in the corresponding non-vulnerable sample (``FIX'' on line 3 in Listing~\ref{lst:nonvul_smp}). The absence of this information in other datasets makes it difficult, both, to identify \textit{VF}s and to assess the accuracy of \textit{VF} detection in samples. Along with the fact that SARD is the largest publicly available vulnerability dataset, containing many real-world security flaws (e.g., from Wireshark and GIMP) and  actively supported by the National Institute of Standards and Technology~\cite{nist}, this led us to direct our focus to the SARD dataset. We will now delve into the process of identifying vulnerability~(\textit{VF}) and spurious~(\textit{SF}) features, followed by the development of the taxonomy.

\begin{minipage}{\linewidth}
  \centering
  \begin{minipage}{0.4825\linewidth}
\begin{lstlisting}[language=C++, 
                caption={Vul. Sample},
                label=lst:vul_smp,
                mathescape,
                % frame=tlrb,
                %escapechar=@,
                xleftmargin=0cm,
                xrightmargin=0cm,
                style = Cpp]
int * data;
data = NULL;
/* FLAW: Allocate memory without using sizeof(int) */
data = (int *)ALLOCA(@10@);
int source[10] = {0};
/* POTENTIAL FLAW: Possible buffer overflow if data was not allocated correctly in the source */
memcpy(data, source, @10*sizeof(int)@);
printIntLine(data[0]);
\end{lstlisting}
 \end{minipage}
  \hspace{0.01\linewidth}
  \begin{minipage}{0.4825\linewidth}
\begin{lstlisting}[language=C++, 
                caption={Non-Vul. Sample},
                label=lst:nonvul_smp,
                mathescape,
                % frame=tlrb,
                %escapechar=@,
                xleftmargin=0cm,
                xrightmargin=0cm,
                style = Cpp]
int * data;
data = NULL;
/* FIX: Allocate memory using sizeof(int) */
data = (int *)ALLOCA(@10*sizeof(int)@);
int source[10] = {0};
/* POTENTIAL FLAW: Possible buffer overflow if data was not allocated correctly in the source */
memcpy(data, source, @10*sizeof(int)@);
printIntLine(data[0]);
\end{lstlisting}
  \end{minipage}
\end{minipage}

\subsection{Identifying Vulnerability Features (\textit{VF})}
\label{subsec:vul_feat}

\begin{table}[b!]
\vspace{-2.25mm}
\centering
\renewcommand{\arraystretch}{1.1}
\addtolength{\tabcolsep}{-2pt}
\caption{Top 10 CWEs in the SARD dataset}
\vspace{-1mm}
\begin{tabular}{lp{6.2cm}l}
\toprule
CWE ID &                                    Description &  \# Files \\
\midrule
CWE805 &      Buffer Access with Incorrect Length Value &             1506 \\
CWE806 &      Buffer Access Using Size of Source Buffer &             1037 \\
CWE124 &                              Buffer Underwrite &              907 \\
CWE127 &                              Buffer Under-read &              784 \\
CWE193 &                               Off-by-one Error &              748 \\
CWE126 &                               Buffer Over-read &              550 \\
CWE415 &                                    Double Free &              421 \\
CWE839 & Numeric Range Comparison Without Minimum Check &              314 \\
CWE131 &           Incorrect Calculation of Buffer Size &              129 \\
CWE416 &                                 Use After Free &              129 \\
\bottomrule
\label{tab:top10cwes}
\end{tabular}
\end{table}

\looseness=-1
We analyzed the annotated descriptions in the SARD dataset to identify properties of code that contribute to a vulnerability. Specifically, we focused on the 10 most frequent vulnerability categories, referred to as CWEs, out of 113 CWEs present in the SARD dataset. These 10 CWEs featured in 6,525 out of SARD's 22,080 source code files, as shown in Table~\ref{tab:top10cwes}. We used as our cut-off point the fact that no other CWEs featured in at least 100 SARD files. 

For each vulnerability in this set, we manually analyzed its annotated descriptions (comments containing prefixes ``FLAW'', ``POTENTIAL FLAW'', or ``FIX'', such as those in Listings \ref{lst:vul_smp} and \ref{lst:nonvul_smp})
and corresponding source code segments. To systematically map the annotated descriptions with the corresponding features in the code, we represent each code sample in a code property graph (CPG)~\cite{yamaguchi2014modeling}. A CPG is constructed from a program's abstract syntax tree (AST), control flow graph (CFG), and program dependence graph (PDG), combining their labeling and property functions. This combination allows a CPG to leverage information from all three sources to conduct better vulnerability analysis~\cite{yamaguchi2014modeling}. 

\begin{figure}[t!]
\centering
\includegraphics[width=0.8\linewidth]{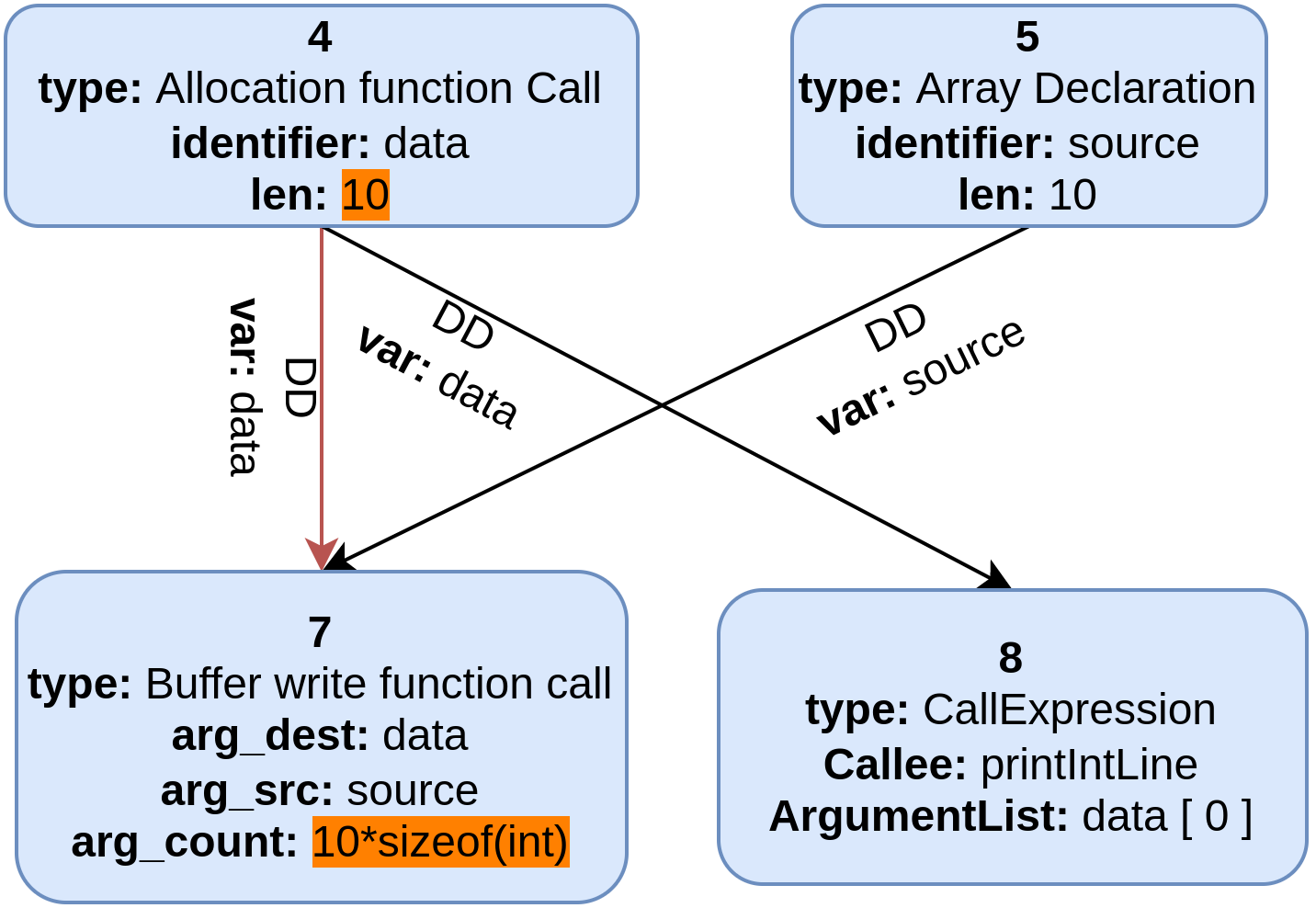}
\captionsetup{width=\linewidth, format=hang}
\caption{Abridged CPG constructed from sample in Listing \ref{lst:vul_smp} 
}
\label{fig:sample_cpg}
\vspace{-3mm}
\end{figure}

For example, Fig.~\ref{fig:sample_cpg} depicts an abridged version of the CPG that is constructed from the sample in Listing~\ref{lst:vul_smp}, where each node represents properties that characterize the statement in the corresponding line of code, and each edge represents data, control, and call dependencies between the nodes. The annotated descriptions in lines 3 and 6 in Listing~\ref{lst:vul_smp} describe the vulnerability present in lines 4 and 7, respectively. They are encoded in the CPG by the leftmost data dependence (DD) edge (in red) for the buffer data between nodes 4 and 7, illustrating an overflow scenario by performing write operation (10*sizeof(int) bytes) on a 10-byte buffer. 
We encoded the description of each of the 10 selected CWE vulnerability categories into a CPG. We elaborate on these rules in Section~\ref{sec:approach}. 

\subsection{Identifying Spurious Features (\textit{SF})}
\label{subsec:spu_feat}

\textit{SF}s required a different approach since all vulnerability datasets focus on  features relevant to vulnerabilities and not on those that should be avoided. As our starting point, we used several existing studies, which provide valuable insights into how features such as variable and method names~\cite{rahman2023towards} and formatting tokens~\cite{risse2023limits} falsely correlate with vulnerabilities. These features are examples of \textit{SF}s we aim to study, and they can have an especially negative impact on the robustness and performance of \emph{token-based} vulnerability detectors, such as SySeVR~\cite{li2021sysevr} and LineVul~\cite{fu2022linevul}. 

To address this, researchers have recently developed \emph{graph-based} vulnerability detectors, such as ReVeal~\cite{chakraborty2021deep} and DeepWukong~\cite{cheng2021deepwukong}, which replace these \textit{SF}s with symbolic names (e.g., \textsc{var1}, \textsc{fun1}}) and incorporate additional information from program graphs (e.g., control and data flow, call dependencies, etc.) to make predictions. This also means that the \textit{SF}s observed in existing literature~\cite{rahman2023towards,risse2023limits}
for token-based detectors do not apply to the graph-based detectors,  requiring further exploration of \textit{SF}s that may influence the latter. 

To this end, we focused on the assumptions made by graph-based detectors~\cite{cheng2021deepwukong, chakraborty2021deep} regarding features that may contribute to a vulnerability. For instance, one common assumption is that a vulnerability is defined strictly by the set of nodes in the graph of the vulnerable sample. However, we observe that the same vulnerability would still exist if a mock node (e.g., a printf("Benign") statement) is added to the graph. Another common assumption is that the set of edges strictly defines the vulnerability in the graph of the vulnerable sample. However, a mock edge (e.g., if (5!=5) return;) can be added to the graph without affecting its vulnerability. The idea of introducing changes that should not impact vulnerability to the sets of nodes and edges was inspired by the existing literature that discussed how graph neural networks learn spurious correlations between sets of nodes and edges~\cite{fan2023generalizing}. Our examination of the SARD dataset revealed that the above two assumptions in particular do not always hold and can lead to spurious correlations that impact detectors' predictions. We demonstrate the extent of this impact in Section~\ref{sec:results}.

      \begin{figure}[b!]
      \vspace{-2mm}
        \begin{forest}
              for tree={
                font=\footnotesize,
                grow'=0,
                child anchor=west,
                parent anchor=south,
                anchor=west,
                calign=first,
                inner ysep=-2pt,
                outer ysep=0pt,
                forked edges,
                edge path={
                    \noexpand\path [draw, \forestoption{edge}]
                    (!u.south west) +(4pt,0) |- (.child anchor) \forestoption{edge label};
                },
                before typesetting nodes={
                    if n=1
                    {insert before={[,phantom]}}
                    {}
                },
                fit=band,
                before computing xy={l=8pt},
              }  
            [
            Code Features
                        [Vulnerability Features
                            [Overflow
                                [Incorrect Calculation of Buffer Size]
                                [Buffer Access Using Size of Source Buffer]
                                [Buffer Access Using Size of Destination Buffer]
                                [Off-by-one Error]
                            ]
                            [Deallocated Buffer Use
                                [Use-After-Free]
                                [Double-Free]
                            ]
                            [Numeric Range Comparison Without Minimum Check
                                [Buffer Underwrite]
                                [Buffer Under-read]
                            ]
                            [Sensitive API Call
                                [Read API]
                                [Write API]
                            ]
                        ]
                        [Spurious Features
                            [Token-based
                                    [\textcolor{gray}{Identifier Names}
                                    ]
                                    [\textcolor{gray}{Code Formatting}
                                    ]
                            ]
                            [Graph-based
                                [Node Set
                                        [Print instructions without variable]
                                        [\textcolor{gray}{Unused variable declarations}]
                                ]
                                [Edge Set
                                    [Unreachable conditional operation without variables]
                                    [\textcolor{gray}{Dependency altering uses of variables}]
                                ]
                            ]
                        ]
            ]
            \end{forest}
          \caption{Taxonomy of Code Features}
          \label{tax:code_feat}
      \end{figure}
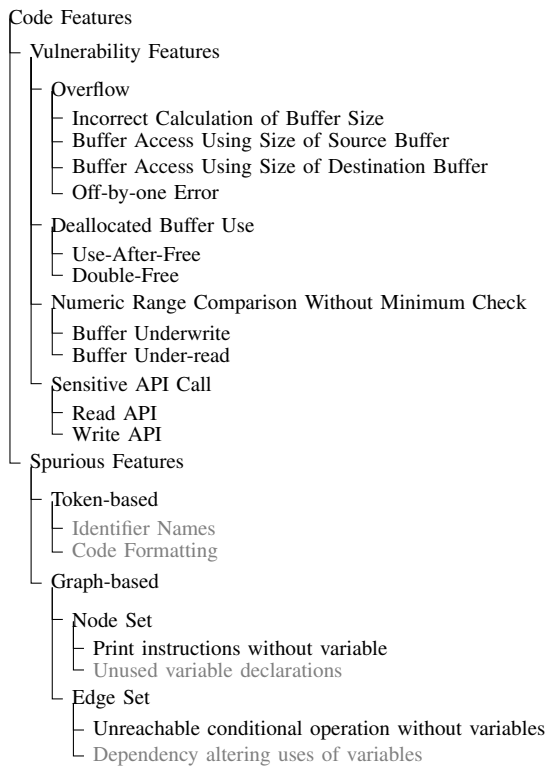

\begin{figure*}[t]
\centering
\includegraphics[width=1\textwidth]{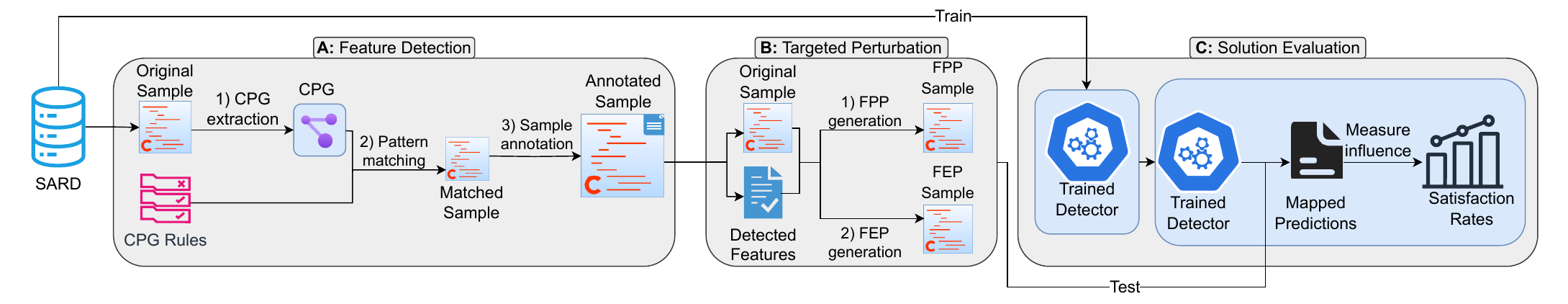}
\vspace{-5mm}
\caption{High-level \textit{VIPer} workflow}
\label{fig:approach}
\vspace{-4mm}
\end{figure*}

\subsection{Developing the Taxonomy}

To capture the dichotomy between \textit{VF}s and \textit{SF}s, and to provide a comprehensive understanding of their characteristics, we classified them into a taxonomy of \textit{Code Features}. 
Our hierarchy-based taxonomy~\cite{usman2017taxonomies}  is shown in Fig.~\ref{tax:code_feat}. 
The taxonomy is not intended to be comprehensive. We expect that follow-on work will add further categories to the taxonomy, which will in turn be encoded as further \textit{VIPer} rules.

\looseness=-1
We partitioned \textit{Code Features} into  two mutually exclusive sub-classes \textit{Vulnerability Features}~(\textit{VF}s) 
and \textit{Spurious Features} (\textit{SF}s). 
All 10 identified \textit{VF}s from Table~\ref{tab:top10cwes} are classified under the \textit{Vulnerability Features} sub-class, which is subdivided into four further sub-classes based on feature characteristics:

\begin{enumerate}
\looseness=-1
\item \textit{Overflow} vulnerabilities occur when a buffer is accessed outside its allocated memory. 
We further categorize overflow vulnerabilities based on the nature of the out-of-bounds access: (a)~Incorrect Calculation of Buffer Size, (b)~Buffer Access Using Size of Source Buffer, (c)~Buffer Access Using Size of Destination Buffer, and (d)~Off-by-one Error.

\item
\textit{Deallocated Buffer Use} vulnerabilities arise when a buffer is accessed after it has been deallocated. We further categorize these vulnerabilities based on the type of post-deallocation use: (a)~Use-After-Free and (b)~Double-Free.

\item
\textit{Numeric Range Comparison} vulnerabilities involve an array indexing operation where the index variable never gets checked for a minimum value. Based on the type of array indexing operation (i.e., read or write), we subdivide these vulnerabilities into (a)~Buffer Underwrite and (b)~Buffer Under-read.

\item
\textit{Sensitive API Call} vulnerabilities are characterized by the use of system APIs that are well-known for causing vulnerabilities \cite{li2018vuldeepecker, cowan1998stackguard, cowan2000buffer}. We categorize these vulnerabilities into (a)~Read APIs and (b)~Write APIs.

\end{enumerate}

We categorized the \textit{Spurious Features} based on the two primary approaches for DL-based vulnerability detection:

\begin{enumerate}
    \item
\textit{Token-based} \textit{SF}s comprise features of individual code tokens that may lead to spurious correlations, such as (a)~Identifier Names that may not be relevant to the vulnerability and (b)~Code Formatting. \textit{SF}s previously identified in literature~\cite{rahman2023towards, risse2023limits} fall under this category (grayed out in Fig.~\ref{tax:code_feat}). 
{
\item 
\textit{Graph-based} \textit{SF}s are features of nodes and their relationships (edges) that exhibit spurious correlations but do not inherently contribute to vulnerabilities. We categorize these features into (a) \textit{Node Set} and (b) \textit{Edge Set}. Node Set refers to the presence of one or more single-line instructions that have no impact on the vulnerability. Specific instances of these \textit{SF}s include (i) \textit{Print instructions without variables}, i.e., ``empty'' print statements in the code, and (ii) \textit{Unused variable declarations}, i.e., variables that are not used after declaration. Edge Set refers to the presence of multi-line instructions defining data or control dependencies that do not affect vulnerability. Specific instances of  Edge Set vulnerabilities are (i) \textit{Unreachable conditional operation without variables}, i.e., a condition that will always be false and may change the control dependency, and (ii) \textit{Dependency-altering uses of variables}, i.e., usage of variables that may alter dataflow or control dependency but that do not contribute to vulnerability. Note that the goal of the study reported in this paper is to observe the impact of \textit{SF} categories that require minimal change. For this reason, we exclude the \textit{SF}s that are greyed out in Fig.~\ref{tax:code_feat} from this study.
}
\end{enumerate}

The transition from individual structured annotations in the SARD dataset to the development of the taxonomy was carried out iteratively. We analyzed the annotated descriptions of each sample containing VFs. During each iteration, an author suggested a category and potential sub-categories, which were then discussed and refined. This process continued until all authors reached a consensus. Additionally, we incorporated categories for SFs by consulting relevant literature~\cite{cheng2021deepwukong,chakraborty2021deep,fan2023generalizing} and applying the same iterative approach.

\section{Approach}
\label{sec:approach}

Our approach to \textbf{\uline{V}}ulnerability \textbf{\uline{I}}dentification via \textbf{\uline{Per}}turbations, \textit{VIPer}, comprises three phases, as illustrated in Fig.~\ref{fig:approach}: (1)~\textit{Feature Detection} uses CPG rules to identify the presence of \textit{VFs} in samples and annotate them as candidates for perturbation. Since \textit{SF}s are inherent properties of code, this phase does not pertain to them. (2)~\textit{Targeted Perturbation} systematically applies FPPs (alterations in code with no change to a feature) and FEPs (alteration in code that eliminates the feature) to the tagged samples, generating a perturbed dataset for each of the 10 \textit{VF}s. For \textit{SF}s, \textit{VIPer} only generates FPPs because \textit{SF}s are inherent properties of code that cannot be eliminated. 
We used the structured annotations in the SARD dataset as ground truth for evaluating the correctness of the CPG rules to detect features. We manually compared the CPG rule-detected features in 363 out of the 6,525 files ($95\%$ confidence with a $5\%$ margin of error) with the annotations in SARD.  When devising the FPPs and FEPs, we ensured that the perturbation targeted the property mentioned in the annotations and later manually analyzed the same 363 files to determine if the FPPs retained the vulnerability and the FEPs removed the vulnerability.  
(3)~\textit{Solution Evaluation} assesses the detectors' responses to the perturbed datasets, examining how perturbations affect the predictive performance of the detectors and their ability to react satisfactorily. A satisfactory reaction is defined as no prediction change for FPPs and a prediction change for FEPs. With this information, \textit{VIPer} quantifies the influence of these features on the detectors' predictions. 
The following subsections  detail each phase.

\vspace{-1mm}
\subsection{Feature Detection}
\label{subsec:feat_det}
{
The goal of the \textit{Feature Detection} phase is to systematically identify the suitable candidate samples for perturbation. The whole process is broken down into three main steps: 1) CPG Extraction, 2) Pattern Matching, and 3) Sample Annotation. We elaborate each step next:

\subsubsection{CPG extraction}
We use joern~\cite{yamaguchi2014modeling} to construct the CPG for each sample in the dataset. To facilitate our analysis, we abridge the graph by retaining only edges that represent the most relevant relationships: Data Dependence (DD), Control Dependence (CD), Post-Dominance (PD), DEF, and USE. These edge types are selected because they are widely recognized as essential relationships for vulnerability analysis~\cite{chakraborty2021deep}. 

\looseness-1
A CPG is formally defined as a graph $\displaystyle G=( V,E,\mu )$, where $V$ is the set of all nodes and $E$  the set of all edges in the CPG, and $\mu :\ ( V\cup E) \times K\rightarrow S$ is a  function that sets or retrieves the property of a node or edge. $K$ is the key that denotes which property would be retrieved or set, and $S$ is the value for property $K$. To facilitate our implementation of \textit{VIPer}, we extended the property function $\mu$'s capabilities to retrieve and set additional properties of nodes and edges. Due to space limitation, we list these additional property keys and values (as acronyms) used by property function $\mu$ in the online appendix~\cite{appendix}.


\subsubsection{Pattern matching}
Once the abridged CPGs are constructed from the dataset samples, we iterate through each CPG to identify which ones satisfy any of our predefined CPG rules for the \textit{VFs}. If a CPG satisfies a rule, we infer that the corresponding dataset sample exhibits the \textit{VF} associated with that rule and is a suitable candidate for perturbation. The CPG rules along with their corresponding \textit{VF} are given in Table~\ref{tab:cpg_rules}. 

To assess if a CPG satisfies a rule, we develop corresponding detection algorithms for each of the rules listed in Table~~\ref{tab:cpg_rules}. Due to the page limitation, we will only describe the detection algorithm for Rule ID: 2.1 for \textit{VF} \textit{Incorrect Calculation of Buffer Size} which is the CPG rule that the sample in Listing~\ref{lst:vul_smp} satisfies. We provide the algorithms for the remaining detection algorithms in the online appendix~\cite{appendix}.


\begin{table*}[htbp]
\centering
\caption{CPG Rules}
\vspace{-1mm}
\begin{tabular}{lm{0.2\linewidth}m{0.7\linewidth}}
\toprule
\textbf{Rule ID} & \textbf{Vulnerability Feature} &\textbf{CPG Rule (condition under which vulnerability exists in a sample)} \\
\midrule 
\midrule 
2.1 & Incorrect Calculation of Buffer Size (IBS) & CPG constructed from the sample must have a node $v$ representing a buffer write function call that writes $n$ number of bytes to a buffer $d$ with a defined length of $LEN_{d}$ that is smaller than $n$ (i.e., $LEN_{d} \ < \ n$). \\
\hline 
2.2 & Buffer Access Using Size of Source Buffer (BSB) & CPG constructed from the sample must have a node $v$ representing a buffer write function call that writes $n$ number of bytes from a source buffer $s$ with a defined length of $LEN_{s}$ equaling $n$ to a destination buffer $d$ with a defined length of $LEN_{d}$ that is smaller than $n$ (i.e., $( LEN_{d} \ < \ n) \ \land \ ( n\ ==\ LEN_{s})$). \\
\hline 
2.3 & Off-by-one Error (OE) & 
CPG constructed from the sample must have a node $v$ representing a buffer write function call that writes $n$ number of bytes to a buffer $d$ with a defined length of $LEN_{d}$ that is smaller than $n$ by exactly one. (i.e., $n\ ==\ LEN_{d} \ +\ 1$). \\
\hline 
2.4 & 
Buffer Over-read (BO) & CPG constructed from the sample must have a node $v$ representing a buffer copy function call that reads $n$ number of bytes from a buffer $s$ with a defined length of $LEN_{s}$ that is smaller than $n$ (i.e., $LEN_{s} \ < \ n$). \\
\hline 
2.5 & Double-Free (DF) & CPG constructed from the sample must have two nodes $u$ and $v$ who call \texttt{free} on the same buffer $b$ and there exists no node $w$ between $u$ and $v$ that uses an allocation function (e.g., \texttt{malloc}) on $b$. \\
\hline 
2.6 & Use-After-Free (UAF) & CPG constructed from the sample must have two nodes $u$ and $v$ where node $v$ uses a buffer $b$ after node $u$ already deallocates buffer $b$. \\
\hline 
2.7 & Buffer Underwrite (BUW) & CPG constructed from the sample must have a node $v$ that writes to a buffer $b$ using an index value $idx$ where $idx$ is never checked to ensure that it does not hold a negative value. \\
\hline 
2.8 & Buffer Under-read (BUR) & CPG constructed from the sample must have a node $v$ that reads from a buffer $b$ using an index value $idx$ where $idx$ is never checked to ensure that it is not a negative number. \\
\hline 
2.9 & Read API (RA) & CPG constructed from the sample must have a node $v$ representing a function call to a sensitive Read API (e.g., \texttt{fgets}) where the location $\mathcal{L}$ of node $v$ is a vulnerable line in the sample. \\
\hline 
2.10 & Write API (WA) & CPG constructed from the sample must have a node $v$ representing a function call to a sensitive Write API (e.g., \texttt{memcpy}) where the location $\mathcal{L}$ of node $v$ is a vulnerable line in the sample. \\
\bottomrule
\end{tabular}

\label{tab:cpg_rules}

\vspace{-4.75mm}
        
\end{table*}

We will now discuss the algorithm used to check if a CPG satisfies the CPG rule with ID 2.1 - \textit{Incorrect Calculation of Buffer Size}. The description for its rule in Table~\ref{tab:cpg_rules} states that, for this vulnerability feature to exist in a sample, the CPG constructed from the sample must have a node $v$ representing a buffer write function call that writes $n$  number of bytes to a buffer $d$ with a defined length of $LEN_{d}$ that is smaller than $n$ (i.e., $LEN_{d} < n$). To detect this feature, \textit{VIPer} checks if the constructed CPG has a node $v$ representing a buffer write function call where the number of bytes to write ($n$) is larger than the defined length of the destination buffer ($LEN_{d}$) by traversing the constructed CPG to determine the static values for $LEN_{d}$ and $n$. Algorithm~\ref{algo:alg2.1} describes how \textit{VIPer} determines these static values and checks if the sample satisfies Rule 2.1.

\begin{algorithm}[htbp]
\small
\DontPrintSemicolon
\KwIn{$G=( V,E, \mu )$ representing the CPG constructed from the sample code}
\KwOut{Boolean value indicating whether the feature exists in the sample\\
\textbf{Property keys for $\mu$}
\begin{itemize}
    \item arg\_dest: Destination buffer in a WF
    \item arg\_count: Number of bytes to write in a WF
    \item type: Type of Node or Edge
    \item len: Length of a defined buffer
    \item var: Variable associated with data dependence
\end{itemize}
\textbf{Begin}}
Let $V' \subset V$ $\gets$ \{$v$ for $v \in V$ if $\mu(v, type)$ = WF\}\;
\For{node $v$ in $V'$}
{
    Let $d$ $\gets$ $\mu(v, arg\_dest)$\;
    Let $n$ $\gets$ $\mu(v, arg\_count)$\;
    Let $IN_{v}^{DD} \subset V$ $\gets$ \{$u$ for $u \in V$ if $(u, v) \in E$ and $\mu(u, v, type) = $ DD and $\mu(u,v, var) = d$ \}\;
    \For{$u$ in $IN_{v}^{DD}$}
    {
        \uIf{$\mu(u, type)$ = AF or $\mu(u, type)$ = AD}
        {
            Let $LEN_{d}$ $\gets$ $\mu(u, len)$\;
            \uIf{$n > LEN_{d}$}
            {
                \Return{true}\;
            }
        }
    }
}
\Return{false}\;
    
\caption{
Incorrect Calculation of Buffer Size}
\label{algo:alg2.1}
\end{algorithm}
\setlength{\textfloatsep}{0pt}

In Algorithm~\ref{algo:alg2.1}, Line 2 iterates through every node $v$ in $V'$ where $V'$ is the set of nodes representing a buffer write function (WF) call (e.g., a call to \texttt{memcpy}) (Line 1). Lines 3 and 4 in the algorithm retrieve the destination buffer ($d$) and the number of bytes to write ($n$) in the function call, respectively. Line 5 retrieves the start nodes of all the incoming data dependence (DD) edges of node $v$ with respect to the destination buffer $d$ (i.e., $IN_{v}^{DD}$). Lines 6-10 iterate through the retrieved data dependence edges and check if an edge's start node $u$ represents an allocation function (AF) or an array declaration (AD). If so, it further checks whether the number of bytes to write ($n$) according to node $v$ exceeds the length defined at node $u$ (Line 9). If true, it concludes that the sample contains the vulnerability \textit{Incorrect Calculation of Buffer Size}.
We do not develop separate detection algorithms for \textit{SF}s because the purpose of these algorithms is to identify relevant samples for perturbation, which is not necessary for \textit{SF}s. Unlike \textit{VF}s, \textit{SF}s (including the ones we focused on i.e., Identifier Names, Code Formatting, Set of Nodes, and Set of Edges) are inherent properties of every sample.

\subsubsection{Sample annotation}
Once a feature is detected in a sample, \textit{VIPer} annotates the sample with the detected feature, relevant line numbers, and variable names, which differ across the 10 \textit{VF}s. For example, when \textit{VIPer} detects \textit{Incorrect Calculation of Buffer Size} in a sample (like in Listing~\ref{lst:vul_smp}), it first lists the name of the detected feature as \textit{Incorrect Calculation of Buffer Size} and lists the line number that defines $LEN_{d}$ (line 4 in Listing~\ref{lst:vul_smp}) and the line number where the value of $n$ is defined (line 7 in Listing~\ref{lst:vul_smp}) and the variable name used for the destination buffer $d$ (\texttt{data} for Listing\ref{lst:vul_smp}). We provide the annotations used for the remaining \textit{VF}s in the online appendix~\cite{appendix}.
After completing the sample annotations, \textit{VIPer} creates a dataset comprising only the annotated samples. This annotated dataset is utilized by the subsequent phase of \textit{VIPer}, \textit{Targeted Perturbation}, which we will describe next.

\subsection{Targeted Perturbation}\label{subsec:TargetPert}

The goal of this phase is to understand how robust the models are to changing input and whether the detectors are able to learn from the \textit{VF}s instead of the SFs. We posit that the detectors should be able to correctly predict the outcome solely based on the presence of \textit{VF} in the code sample and should remain unchanged if the code sample changes without losing the \textit{VF}. To achieve this, \textit{VIPer} perturbs samples in two ways: preserving the feature (FPP) or eliminating it (FEP). The underlying principle is that if a perturbation leaves the \textit{VF} intact (FPP), the detector's prediction should remain unchanged. Conversely, if a perturbation eliminates the \textit{VF} (FEP), the detector's prediction should change accordingly.

Fig.~\ref{fig:approach} provides a high-level overview of how \textit{VIPer} generates perturbations. The input for this phase is the annotated dataset generated by the previous phase, \textit{Feature Detection}. From this dataset, \textit{VIPer} extracts the three essential elements required for perturbation: (1) the original sample in the dataset where the \textit{VF} was detected, (2) the name of the detected \textit{VF} and (3) the relevant line numbers and variable names for generating perturbations. Based on the detected feature's name, \textit{VIPer} selects one of 10 tailored perturbation generation algorithm sets. Each set involves generating one or more FPPs and FEPs. Due to page limitations, we will only elaborate on the perturbation generation algorithm set for \textit{Incorrect Calculation of Buffer Size}. The remaining nine algorithm sets are provided in the online appendix~\cite{appendix}. Recall the CPG rule for \textit{VF} \textit{Incorrect Calculation of Buffer Size} is $LEN_{d} < n$. The algorithms used for generating FPPs and FEPs for this CPG rule are given as follows:

\subsubsection{FPP generation}
\looseness=-1
When generating FPPs, we want to make sure that the perturbation still retains the rule $LEN_{d} < n$. There are two ways this can be achieved, (1) decreasing the value of $LEN_{d}$ or (2) increasing the value of $n$. To apply these two types of perturbations, \textit{VIPer} uses Algorithm~\ref{algo:fpp_alg2.1}. In lines 1 \& 2 of the algorithm, it extracts the values of $LEN_{d}$ and $n$ from line numbers extracted from the feature annotated samples. In lines 3 \& 5, it creates two clones of the original sample, and in the first clone applies perturbation (1) (i.e., decreasing the value of $LEN_{d}$) (see line 4) while in the second clone, it applies perturbation (2) (i.e., increasing the value of $n$) (see line 6).

\begin{algorithm}[t]
\small
\DontPrintSemicolon
\KwIn{\\
$G$ = CPG constructed from the sample code\\
$feat\_name$ = name of the detected feature\\
$u$ = line defining the destination buffer\\
$v$ = line representing WF
}
\KwOut{Boolean value indicating whether the feature exists in the sample\\
\textbf{Begin}}
Let $LEN_{d}$ $\gets$ $\mu(u, len)$\;
Let $n$ $\gets$ $\mu(v, arg\_count)$\;
Let $G_{1}=( V_{1},E_{1}, \mu_{1} )$ $\gets$ G.clone()\;
$\mu_{1}(u, len)$ $\gets$ $LEN_{d} - 1$ \;
Let $G_{2}=( V_{2},E_{2}, \mu_{2} )$ $\gets$ G.clone()\;
$\mu_{2}(v, arg\_count)$ $\gets$ $n + 1$ \;
\Return{$G_{1}$, $G_{2}$}\;
    
\caption{
FPP: Incorrect Calc. of Buffer Size}
\label{algo:fpp_alg2.1}
\end{algorithm}

\subsubsection{FEP generation}
When generating FEPs, we want to achieve the opposite goal and ensure that the perturbation no longer satisfies the rule $LEN_{d} < n$. Again, there are two ways this can be achieved, (1) increasing the value of $LEN_{d}$ to match the value of $n$ or (2) decreasing the value of $n$ to match the value of $LEN_{d}$. To apply these two types of perturbations, \textit{VIPer} uses Algorithm~\ref{algo:fep_alg2.1}. Similar to the previous algorithm, in lines 1 \& 2 of Algorithm~\ref{algo:fep_alg2.1}, \textit{VIPer} extracts the values of $LEN_{d}$ and $n$ from line numbers extracted from the feature annotated samples and in lines 3 \& 5, it creates two clones of the original sample. However, unlike the previous algorithm, to generate FEPs, \textit{VIPer} increases the value of $LEN_{d}$ to match $n$ for the first clone (see line 4), and for the second clone, it decreases the value of $n$ to match $LEN_{d}$ (see line 6).

\begin{algorithm}[t]
\small
\DontPrintSemicolon
\KwIn{\\
$G$ = CPG constructed from the sample code\\
$feat\_name$ = name of the detected feature\\
$u$ = line defining the destination buffer\\
$v$ = line representing WF
}
\KwOut{Boolean value indicating whether the feature exists in the sample\\
\textbf{Begin}}
Let $LEN_{d}$ $\gets$ $\mu(u, len)$\;
Let $n$ $\gets$ $\mu(v, arg\_count)$\;
Let $G_{1}=( V_{1},E_{1}, \mu_{1} )$ $\gets$ G.clone()\;
$\mu_{1}(u, len)$ $\gets$ $n$ \;
Let $G_{2}=( V_{2},E_{2}, \mu_{2} )$ $\gets$ G.clone()\;
$\mu_{2}(v, arg\_count)$ $\gets$ $LEN_{d}$ \;
\Return{$G_{1}$, $G_{2}$}\;
    
\caption{
FEP: Incorrect Calc. of Buffer Size}
\label{algo:fep_alg2.1}
\end{algorithm}


To generate perturbations targeting Spurious Features (SFs) in vulnerable samples, we employ separate approaches for token-based and graph-based detectors. Since \textit{VIPer} evaluates token-based approaches for \textit{SF}s that are established in previous literature, we use existing methods of perturbation for these \textit{SF}s. Specifically, we adopt the symbolization mechanism from \textit{Li et al.}~\cite{li2018vuldeepecker} to perturb identifier names and leverage the auto-indentation feature of the CLion IDE\cite{clion}  to introduce indentations into code samples. 

\looseness-1
In contrast, for graph-based \textit{SF}s, we develop new perturbations. Importantly, since modifications to \textit{SF}s do not impact the sample's vulnerability ground truth, all generated perturbations for \textit{SF}s are FPPs. The goal for \textit{SF} perturbations is to modify the nodes and edges in a sample with minimal possible change, without affecting the vulnerability ground truth. Therefore, for \textit{Node Set},
the corresponding perturbation is inserting a \textit{printf("");} statement at the start of each function (i.e., the \textit{Print instructions without variables SF})
and for \textit{Edge Set}, the corresponding perturbation is inserting \textit{if(0==1) return;} at the start of each function (i.e., the \textit{Unreachable conditional operation without variables SF}). \footnote{For better readability, we are referring to the \textit{SF}s \textit{Print instructions without variables} and \textit{Unreachable conditional operation without variables} as \textit{Node Set} and \textit{Edge set}, respectively, for the rest of the paper.}
Note that these perturbations are generic and are applied to all samples in the dataset.

\subsection{Solution Evaluation}\label{subsec:svdt_eval}

The goal of this phase is to measure the effect of the perturbations generated in the previous phase on the predictions of the detectors. By analyzing the predictions of the detectors on the perturbed dataset, \textit{VIPer} measures how the \textit{VF}s and \textit{SF}s influence the detector's prediction. Fig.~\ref{fig:approach} depicts a high-level overview of the \textit{Solution Evaluation} phase. First, we train the detectors on the SARD dataset. Next, we use the dataset samples annotated in the \textit{Feature Detection} phase and their corresponding FPPs and FEPs generated in the \textit{Targeted Perturbation} phase to retrieve the predictions of the detectors. Using these predictions, \textit{VIPer} analyzes the detector's response to the perturbations. Specifically, it measures how satisfactory are the detectors' responses. Recall that a satisfactory reaction is when a detector retains its predictions for FPP perturbations or when it changes predictions for FEP perturbations. Based on this information, we calculate the satisfaction rate $SR_{f}$  of the detectors on the perturbations targeting a feature $\displaystyle f$ for a dataset $\displaystyle X$ to measure the influence of $\displaystyle f$ on the detector's prediction. Formally, the satisfaction rate is defined as:

\begin{equation*}
SR_{f}=\left(\frac{T_{FPP} '+T_{FEP} '}{T_{FPP} +T_{FEP}}\right) \times 100
\end{equation*}
\looseness=-1
where $\displaystyle T_{FPP}$ is the total number of FPPs generated from $\displaystyle X$ for $f$, $\displaystyle T_{FEP}$ is the total number of FEPs generated from $\displaystyle X$ for $f$, $\displaystyle T_{FPP} '$ is the total number of FPPs that retain the detector's prediction (expected outcome), $\displaystyle T_{FEP} '$ is the total number of FEPs that change the prediction of the detector (expected outcome).

Additionally, the aim is to understand how FPPs and FEPs individually impact the detectors. \textit{VIPer} measures this impact by calculating the satisfaction rate of FPPs $SR^{FPP}_{f}$ and FEPs $SR^{FEP}_{f}$ for a feature $f$ individually as shown below:
\begin{equation*}
SR^{FPP}_{f}=\left(\frac{T_{FPP}' }{T_{FPP}}\right) \times 100 ~~~~~~
SR^{FEP}_{f}=\left(\frac{T_{FEP}'}{T_{FEP}}\right) \times 100
\end{equation*}

\section{Evaluation Setup}
\label{sec:exp}

\subsection{Research Questions}

Our evaluation aims to answer three research questions.

\begin{itemize}

\item \textbf{\textit{RQ1}} -- How do the targeted perturbations affect the reported prediction accuracy of the vulnerability detectors?


\item \textbf{\textit{RQ2}} -- What are the detectors' responses to perturbations targeting different features?


\item \textbf{\textit{RQ3}} -- To what extent do the FPPs and FEPs influence the detectors' predictions?


\end{itemize}

\subsection{Classifying Evaluation Results}


For the purpose of our analysis and discussion, specifically RQ3, we classify detector responses into one of four  {$SR^{FPP}_{f}$--$SR^{FEP}_{f}$} combinations: high--high (HH), high--low (HL), low--high (LH), and low--low (LL). 
We consider the satisfaction rate high if the value is within $3\%$ (chosen value $\epsilon$) of the average satisfaction rate for a perturbation type targeting feature $f$. This ensures that detectors performing close to the average are still recognized as effective.\footnote{
While our analysis would be carried out the same way regardless of the selected value $\epsilon$, we selected this  value based on established conventions, as an $\epsilon$ of $0.01 - 0.03$ is frequently used in surveys, such as the American Statistical Association~\cite{asa} and the American Community Survey’s methodology for population and housing estimates~\cite{acs}.
}
Detectors in the \emph{HH} category exhibit a desired understanding of feature $f$ since they retain predictions for FPPs and change predictions for FEPs. These detectors will not experience significant drop in precision or recall in the presence of perturbations. 
A detector in the \emph{HL} category exhibits an understanding of feature $f$, but that understanding does not align with the ground-truth characteristics of $f$ (i.e., the relevant CPG rule). These detectors will experience a noticeable drop in precision in the presence of perturbations. 
A detector in the \emph{LH} category changes its predictions given any perturbation involving feature $f$. Such detectors will experience a significant drop in recall. 
Lastly, detectors in the \emph{LL} category will exhibit a significant drop in both precision and recall in the presence of perturbations.


{Researchers and engineers can use \textit{VIPer}'s results to assess the vulnerability detectors and decide which option is best suited for their purpose. Ideally, one would always prefer a detector falling under the HH category. If none of the available detectors fall under that category, their users will have to assess whether a detector that falls within one of the other categories is suitable for their tasks. Generally, vulnerability detection tasks prioritize recall over precision, i.e., tolerating false alarms while fixing as many vulnerabilities as possible. In such cases, the engineer may prefer detectors falling under the HL category. However, there may be certain tasks where precision is preferred over recall. An example is as a large and stable system in which vulnerable code is less prevalent and the cost of handling false alarms is significant. In that case, the LH category may be considered. } 


\subsection{Evaluation Subjects}

We investigated whether five state-of-the-art vulnerability detectors learn from \textit{VF}s or \textit{SF}s: DeepWukong~\cite{cheng2021deepwukong}, Reveal\cite{chakraborty2021deep}, DeepDFA~\cite{steenhoek2024dataflow}, LineVul~\cite{fu2022linevul}, and SySeVR~\cite{li2021sysevr}.

\textit{DeepWukong} leverages an advanced graph neural network (GNN) to embed code fragments into a low-dimensional representation~\cite{cheng2021deepwukong}. 

\textit{ReVeal} generates function-level prediction by using gated GNNs that are intended to make the model capable of understanding complex code semantics and dependencies~\cite{chakraborty2021deep}.

\textit{DeepDFA} detects function-level vulnerabilities by extracting abstract dataflow information from functions and applying a Gated Graph Sequence Neural Network to learn vulnerabilities in the extracted dataflow~\cite{steenhoek2024dataflow}.

\textit{LineVul} predicts software vulnerabilities at the line level leveraging the BERT architecture~\cite{devlin2018bert} with self-attention layers. It first predicts the vulnerable functions and then locates the specific vulnerable lines within those functions.

\textit{SySeVR} uses syntax-based vulnerability candidates (SyVCs) from code and semantic-based candidates (SeVCs) from control and data dependencies, by representing them into vectors suitable for marking the vulnerabilities in code~\cite{li2021sysevr}.


\subsection{Evaluation Dataset}

\looseness-1
To assess the five evaluation subjects, \textit{VIPer} uses the largest vulnerability dataset, SARD, containing production, synthetic, and academic programs (a.k.a. test cases). We utilized the curated version of SARD from SySeVR as it is the largest dataset used among all five vulnerability detectors~\cite{syse_sard}.
The dataset includes 22,080 C/C++ files. The vulnerable programs in SARD provide precise locations of each vulnerability, enabling effective analysis. In total, the dataset had 366,419 C/C++ functions that we used for training the five evaluation subjects. 



\section{Results}\label{sec:results}


In this section, we present the results corresponding to the three posed research questions in the previous section.

\subsection{RQ1: Analysis of Accuracy} 

\looseness=-1
\emph{RQ1} investigates the impact of targeted perturbations on the detectors' predictive performance. We assess the detectors' performance using two widely adopted evaluation metrics~\cite{powers2020evaluation} - Precision and Recall. Table \ref{tab:metric-change} presents the changes between the original and the perturbed dataset for each identified \textit{VF} and \textit{SF} in terms of predictive performance. A negative value indicates a decline, whereas a positive value indicates an increase in the corresponding metric.
For perturbations targeting \textit{VF}s that belong to \textit{Overflow} vulnerability sub-class in Fig.~\ref{tax:code_feat}, overall, we observe that all five detectors suffer a noticeable decrease in precision. However in case of SySeVR, the drop is $48\%$ on average, higher than the other four detectors. As for recall, DeepWukong shows a drop up to $25.78\%$ while the rest of the detectors show minimal drop with LineVul retaining its original recall.

\looseness-1
For \textit{VF}s that belong to \textit{Deallocated Buffer Use} vulnerability sub-class, we observe that all graph-based detectors (DeepWukong, ReVeal, and DeepDFA) noticeably outperform the token-based detectors (LineVul and SySeVR). DeepWukong achieves the smallest drop in precision ($3.3\%$ on average) and a slight increase in recall ($1.99\%$ on average), while DeepDFA exhibits the highest drop in precision (up to $24\%$). Since both token-based approaches demonstrate relatively poor performance (i.e., over $50\%$ drop in precision on average for LineVul and similar drop in recall for SySeVR), this could be due to the fact that \textit{Deallocated Buffer Use} vulnerabilities are characterized mainly by the control flow of a program ~\cite{yamaguchi2014modeling}. Since graph-based detectors, by design, are able to better capture context information from graphs, they are expected to outperform token-based approaches. However, DeepDFA only focuses on the dataflow of a function which may explain its drop in precision.
For \textit{VF}s that belong to \textit{Numeric Range Comparison} vulnerabilities, LineVul shows the highest drop in precision ($33\%$) among the five detectors.

\begin{table*}[t!]
\centering
\renewcommand{\arraystretch}{1.05}
\caption{Accuracy Metrics Change (Original$\rightarrow$ Perturbed) [least values are in bold]}
\vspace{-1mm}
\begin{tabular}{ll|rrrrrr|rrrr}
\toprule
&  & \multicolumn{6}{c|}{Graph-based} & \multicolumn{4}{c}{Token-based}  \\
& Feature Name & \multicolumn{2}{c}{DeepWukong} &  \multicolumn{2}{c}{ReVeal} &  \multicolumn{2}{c|}{DeepDFA}  &   \multicolumn{2}{c}{LineVul} & \multicolumn{2}{c}{SySeVR} \\
& & PREC & REC & PREC & REC & PREC & REC  &  PREC & REC &  PREC & REC \\
\midrule
\midrule
\multirow{10}{*}{\rotatebox[origin=c]{90}{\textit{VF}}} 
& IBS &    -33.38 & \textbf{-9.55}  &    -35.42 & 0.00  &    -9.94 & -2.57    &   -35.42 & 0.00  &    \textbf{-49.70} & 0.00  \\
& BSB &   -52.40 & \textbf{-25.78}  &    -54.39 & 0.00  &     -8.31 & -0.76    &       -54.39 & 0.00  &   \textbf{-69.70} & 0.00  \\
& OE &       -12.78 & 4.87  &    -24.55 & -1.97  &             -7.71 & -4.20      &     -21.94 & 0.00  &     \textbf{-32.33} & \textbf{-5.95}  \\
& BO &       -11.85 & 4.31  &   -22.71 & \textbf{-1.21}  &    -14.40 & -6.60   &     -22.53 & 0.00  &      \textbf{-41.76} & 2.72  \\
& DF &         0.00 & 0.00  &    -18.71 & -4.04  &   -24.23 & -1.38      &   \textbf{-60.00} & 0.00  &        0.00 & \textbf{-57.71}  \\
& UAF &         -6.69 & 3.99  &   -18.65 & -3.42  &    -23.02 & -1.31      &     \textbf{-47.25} & 0.00  &      1.29 & \textbf{-51.05}  \\
& BUW &  -34.02 & \textbf{-16.27}  &    -9.41 & -4.84  &   -10.94 & -3.39     &   \textbf{-45.88} & -1.43  &       -30.43 & 1.38  \\
& BUR &    -10.45 & \textbf{-6.69}  &    -22.13 & -6.12 &   -21.44 & -4.81     &    -20.15 & -1.24  &     \textbf{-28.01} & -3.44  \\
& RA &   0.00 & 0.00  &  \textbf{-100.00} & 0.00 &    9.92 & 1.66        &  \textbf{-100.00} & 0.00  &  \textbf{-100.00} & 0.00  \\
& WA &     -30.48 & \textbf{-4.61}  &       \textbf{-57.04} & 0.01  &    1.51 & 0.47                  &   \textbf{-57.04} & 0.01  &    -29.99 & 0.95  \\
\midrule
\multirow{4}{*}{\rotatebox[origin=c]{90}{\textit{SF}}} 
& Identifier Name &    - & -  &     - & -  &   - & - &     -67.69 & -0.35  &       -6.70 & \textbf{-10.41}  \\
& Code Formatting &     - & -  &      - & -   &  - & - &   -0.41 & -0.03  &        -0.01 & 0.00  \\
& Node Set &     \textbf{-3.90} & -0.74  &   -3.44 & \textbf{-18.42}  &   0.16 & 0.03    &    - & -  &         - & -  \\
& Edge Set &     -1.08 & \textbf{-29.04}  &  \textbf{-3.87} & -21.38  &   0.06 & -0.01   &    - & -  &         - & - \\
\midrule
\bottomrule
\label{tab:metric-change}
\end{tabular}
\vspace{-8mm}
\end{table*}

For \textit{Sensitive API Call} \textit{VF}s, LineVul and ReVeal show the biggest drop in precision by $78\%$ on average with LineVul also showing the largest drop in recall ($50\%$). SySeVR gets the second-largest drop in precision with $65\%$ on average.  Results for \textit{VF}s under this sub-class are surprising since this \textit{VF} focuses on calls to specific system APIs; token-based approaches are by design expected to effectively leverage tokens containing system API names. We elaborate on this particular aspect in Section~\ref{sec:discuss}.

\looseness-1
For the \textit{SF} \textit{Code Formatting}, we observe that both SySeVR and LineVul are reasonably resilient with the drop in precision and recall never exceeding $1\%$. 
For the \textit{SF} \textit{Identifier Name}, LineVul shows a significant drop in precision with over $67\%$, while SySeVR's precision drops only by $6.7\%$. This is likely because SySeVR preprocesses raw code tokens (e.g., symbolizing code elements to prevent learning from custom naming conventions) while LineVul converts the raw code tokens directly into vectors thus exposing itself to this \textit{SF}.
For two \textit{SF}s that belong to the graph-based sub-class, we observe that DeepWukong only suffers a noticeable drop in recall for the \textit{SF} \textit{Edge Set} ($29.04\%$), while ReVeal's recall drops for both \textit{SF}s. This reduction in recall in both cases suggests that graph-based approaches tend to generate false negatives whenever encountering a behavior-preserving perturbation, such as introducing a mock edge to the graph. However, in case of \textit{Node Set} perturbations, DeepWukong is less influenced, possibly due to the fact that it only considers control and data dependence edges from the CPG; since \textit{Node Set} perturbations do not introduce any of these dependencies, they do not influence DeepWukong's predictions. In contrast, ReVeal considers all the nodes from CPG, and is thus inherently exposed to \textit{Node Set} perturbations.
The precision and recall drops for DeepDFA are negligible ($<0.5\%$). Similarly to the above discussion of DeepWukong, this is due to the fact that DeepDFA only works on dataflow information and \textit{SF} perturbations introduce no changes to a program's dataflow. 
At the same time, it may be worth investigating whether DeepDFA retains the same robustness against \textit{SFs} that impact dataflow information, e.g., the use of intermediate variables during a  mathematical calculation. 

\subsection{RQ2: Analysis of Feature Satisfaction Rate}
RQ2 investigates the detectors' robustness to perturbations targeting individual features using the $SR_{f}$ metric introduced in Section~\ref{subsec:svdt_eval}. Fig.~\ref{fig:heatmap} presents a heatmap illustrating the $SR_{f}$ of the detectors for feature-specific perturbations -- The larger the $SR_{f}$, the more intense the color in the cell.
Among the five detectors, DeepWukong has the highest average $SR_{f}$ for perturbations targeting \textit{VF}s that belong to \textit{Overflow} vulnerability sub-class with $88\%$ while SySeVR has the lowest average $SR_{f}$ with $61\%$.
For \textit{VF}s that belong to \textit{Deallocated Buffer Use} vulnerability sub-class, we observe an opposing scenario from RQ1 where both token-based detectors demonstrate a better $SR_{f}$ ($15\%$ higher) than the graph-based detectors.


For \textit{VF}s that belong to \textit{Numeric Range Comparison} vulnerability sub-class, SySeVR displays the lowest $SR_{f}$ with $57\%$ on average while rest of the detectors demonstrate higher $SR_{f}$, i.e., $86\%-87\%$.
For \textit{Sensitive API Call} \textit{VF}s, DeepWukong achieves the highest $SR_{f}$ with $95\%$ on average. It is worth noting that, for the feature \textit{Read API}, \textit{VIPer} does not generate FPPs since we could not find replacement for system read functions from the list of vulnerability-causing system APIs~\cite{li2018vuldeepecker, cowan1998stackguard, cowan2000buffer} with matching argument list and syntax. Therefore, \textit{VIPer} only generates FEPs for this \textit{VF}. Since all the FEPs targeting this feature retain predictions for ReVeal and LineVul, their $SR_{f}$ is $0$. We will discuss this particular detectors' behavior in the next Section~\ref{sec:discuss}.

For \textit{SF}s that belong to the \textit{Token-based} sub-class, both LineVul and SySeVR produce roughly the same $SR_{f}$.
For \textit{SF}s that belong to graph-based sub-class,  DeepWukong achieves a slightly higher $SR_{f}$ 
than ReVeal, while DeepDFA achieves near perfect $SR_{f}$ for both \textit{SFs}.

\setlength{\textfloatsep}{12pt}
\begin{figure}[b!]
\vspace{-4mm}
\centering
\includegraphics[width=1.0\linewidth]{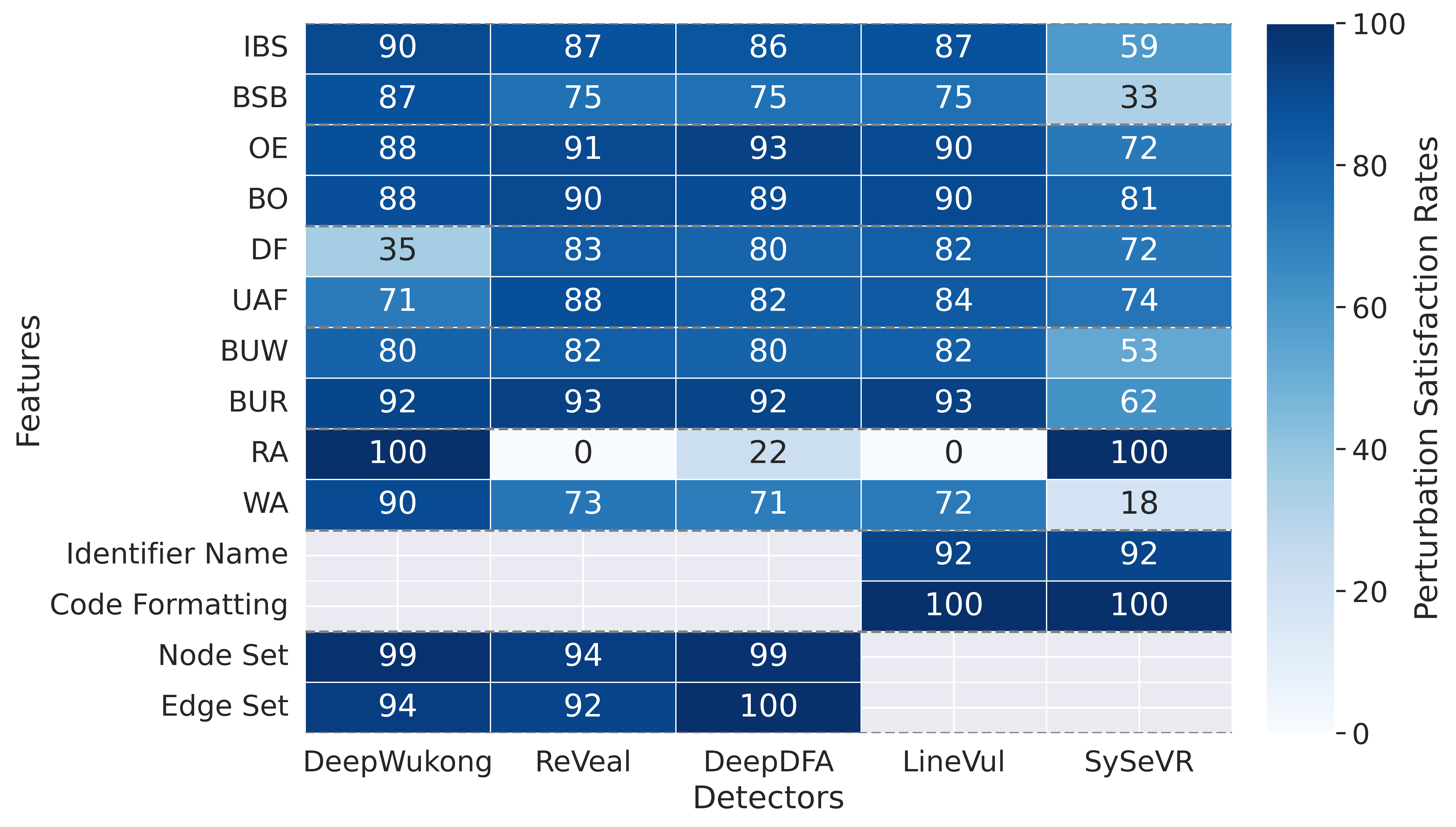}
\captionsetup{width=\linewidth, format=hang}
\caption{Perturbation Satisfaction Rates
}
\label{fig:heatmap}
\end{figure}


\subsection{RQ3: Analysis of FPP and FEP Satisfaction Rate}
The goal of RQ3 is to determine which type of perturbations (FPPs or FEPs) have the most significant impact on the detectors' predictions. Table~\ref{tab:fep-fpp-sat-rate} shows the $SR^{FPP}_{f}$ and $SR^{FEP}_{f}$ for individual \textit{VFs}.
As can be seen in Table~\ref{tab:fep-fpp-sat-rate}, for perturbations targeting \textit{VF}s that belong to \textit{Overflow} vulnerability sub-class, the average $SR^{FPP}_{f}$ is $99.3\%$ and since all five detectors' $SR^{FPP}_{f}$ are around \textasciitilde$3\%$ of the average we consider all their $SR^{FPP}_{f}$ to be high. However, the average $SR^{FEP}_{f}$ is $9.8\%$, which is lower than the recommended minimum threshold for metrics measuring the intelligence of AI systems~\cite{cabitza2019wants} (i.e., $51\%$). Therefore, if the $SR^{FEP}_{f}$ is lower than $51\%$ we consider the value to be low. This means that all five detectors receive very low $SR_{f}$; specifically, LineVul never changes its prediction for FEPs. Therefore, all five of them fall under the category HL.
For \textit{VF}s falling that belong to \textit{Deallocated Buffer Use} vulnerability sub-class, using the same principle as before, we selected thresholds for $SR^{FPP}_{f}$ and $SR^{FEP}_{f}$ as $93.88\%$ and $51\%$ respectively. We observe that DeepDFA, DeepWukong, LineVul, and ReVeal remain in category
HL while SySeVR falls under category LH for deallocated buffer use vulnerabilities.
For \textit{VF}s that belong to \textit{Numeric Range Comparison} vulnerability sub-class, the selected thresholds for $SR^{FPP}_{f}$ and $SR^{FEP}_{f}$ are $97.66\%$ and $51\%$ respectively. Therefore, all five detectors fall under the category HL.
As mentioned when discussing RQ2, \textit{VIPer} does not generate FPPs for \textit{Read API} \textit{VF}, therefore, we only consider $SR^{FEP}_{f}$. We observe that DeepWukong and SySeVR have a $100\%$ $SR^{FEP}_{f}$ while LineVul and ReVeal demonstrate $0\%$ $SR^{FEP}_{f}$, with DeepDFA achieving $22\%$ $SR^{FEP}_{f}$. Note that since we do not have any FPPs for this specific \textit{VF}, we cannot characterize the detectors under one of the four categories.
For Write API \textit{VF}s, the selected thresholds for $SR^{FPP}_{f}$ and $SR^{FEP}_{f}$ are $99.68\%$ and $51\%$ respectively. For this \textit{VF} also, all five detectors fall under the category HL.


\begin{table*}[t!]
\centering
\renewcommand{\arraystretch}{1.15}
\caption{Satisfaction Rates for FPP and FEP}
\vspace{-1mm}
\label{tab:fep-fpp-sat-rate}
\begin{tabular}{l|rrrrrr|rrrr}
\toprule
& \multicolumn{6}{c|}{Graph-based} & \multicolumn{4}{c}{Token-based}  \\
\textit{VF} &      \multicolumn{2}{c}{DeepWukong} & \multicolumn{2}{c}{ReVeal}  &   \multicolumn{2}{c|}{DeepDFA}  &  \multicolumn{2}{c}{LineVul} &   \multicolumn{2}{c}{SySeVR}   \\
& FPP & FEP & FPP & FEP & FPP & FEP & FPP & FEP  & FPP & FEP \\
\midrule
\midrule
IBS &   99.33 & 32.13 & 100.00 & 0.00 & 97.96 & 0.00  & 100.00 & 0.00 &    100.00 & 0.00 \\
BSB &   99.25 & 47.50 & 100.00 & 0.00 & 98.94 & 1.61  & 100.00 & 0.00 &    100.00 & 0.00 \\
OE &    99.63 & 5.30 & 98.18 & 19.51 &  98.20 & 48.84 & 100.00 & 0.00 &    96.92 & 20.63 \\
BO &   99.97 & 10.26 & 99.43 & 4.96 &   99.44 & 0.76  & 100.00 & 0.00 &     95.89 & 17.14 \\
DF &   100.00 & 0.00 & 94.68 & 30.95 &  94.15 & 16.67 & 100.00 & 0.00 &    64.29 & 91.67 \\
UAF &   100.00 & 1.19 & 98.20 & 35.66 & 95.66 & 12.40 & 100.00 & 0.00 &    93.86 & 44.30 \\
BUW &   91.45 & 34.92 & 98.52 & 5.23 &  97.55 & 3.36  & 100.00 & 0.00 &      94.43 & 8.58 \\
BUR &    99.29 & 5.88 & 98.87 & 4.71 &  98.54 & 5.82  & 100.00 & 0.00 &      98.72 & 2.08 \\
RA & - & 100.00 & - & 0.00 &           - & 22.22  & - & 0.00 &  - & 100.00 \\
WA &   98.73 & 43.93 & 100.00 & 0.00 &    97.82 & 2.50   & 100.00 & 0.00 &   100.00 & 18.37 \\
\midrule
\bottomrule
\end{tabular}
\vspace{-4.5mm}
\end{table*}

\section{Discussion and Implications}\label{sec:discuss}



\looseness=-1
The evaluation of five detectors by \textit{VIPer} yielded valuable insights, revealing the detectors' respective strengths and weaknesses.
In the case of DeepWukong, \textit{VIPer} revealed that its lowest $SR_{f}$ comes from \textit{VF}s \textit{Double-Free} and \textit{Use-After-Free}, which both belong to the sub-class \textit{Deallocated Buffer Use}. Given that these vulnerabilities are primarily characterized by program control flow~\cite{yamaguchi2014modeling}, \textit{VIPer}'s findings suggest that DeepWukong struggles to comprehend control dependencies and their contribution to vulnerability. \textit{VIPer} shows that DeepWukong is also influenced by the \textit{SF} \textit{Edge Set}, which may be a key factor underlying its limited understanding of control dependencies. 

\textit{VIPer}'s findings indicate that ReVeal struggles to understand the impact of system APIs on vulnerabilities as it displays the lowest $SR_{f}$ for \textit{VF} \textit{Write API}. ReVeal also shows a drop in recall for \textit{Node Set} and \textit{Edge Set} \textit{SF}s. \textit{VIPer}'s findings suggest that these \textit{SF}s may impede ReVeal's ability to recognize crucial dependencies related to system API calls.

Similarly to ReVeal, DeepDFA struggles to understand how system APIs impact vulnerabilities, evidenced by it demonstrating the lowest $SR_{f}$ for the \textit{VF} \textit{Write API}. This is because DeepDFA does not capture the system APIs in the dataflow analysis during the learning process. On the other hand, unlike ReVeal, DeepDFA is not affected by graph-based \textit{SF}s.

\textit{VIPer} shows that graph-based detectors ReVeal and DeepWukong experience a drop in recall from the \textit{SF} \textit{Edge Set}. This may be attributed to SARD's fix generation process for certain vulnerable code. Specifically, we observed that SARD often fixes vulnerabilities involving sensitive system APIs by introducing control dependencies with unsatisfiable conditions (e.g., \textit{if(5!=5)}). We observed this in 326 source files. \textit{VIPer}'s findings suggest that these control dependencies may lead graph-based detectors to mistakenly associate them with fixes for the vulnerability. There could be two possible remedies to improve the accuracy of vulnerability detection models: 1) we recommend avoiding such spurious control dependencies when generating fixes in synthetic vulnerability datasets, or 2) preventing vulnerability detectors from learning from the spurious control dependencies by eliminating them using an edge filtering algorithm during preprocessing. Since DeepDFA incorporates the above-mentioned remedies by only extracting dataflow information, it does not get influenced by these \textit{SF}s.

\textit{VIPer}'s findings also indicate that LineVul achieves a $100\%$ $SR^{FPP}_{f}$ but a $0\%$ $SR^{FEP}_{f}$. This disparity suggests that LineVul severely overfits when trained on SARD, excelling in false positive reduction but failing to generalize effectively. \textit{VIper} also shows that the \textit{SF}, \textit{Identifier Name} impacts LineVul's predictions by reducing its precision. This \textit{SF} is one of the contributors to LineVul's overfitting to SARD. Therefore, LineVul may benefit from using some preprocessing technique on the tokens (e.g., token symbolizing) to avoid exposing itself to the \textit{SF} \textit{Identifier Name}. For SySeVR, it gets lower $SR_{f}$ compared to the four other detectors for most of the features. However, the lowest value is from the \textit{VF}, \textit{Write API}, suggesting that SySeVR's token symbolization may inadvertently capture system APIs, hindering its ability to learn from these API names and understand their contribution to vulnerabilities. SySeVR may address this by modifying its token symbolization to include only user-defined functions.

\section{Threats to validity}\label{sec:threats}

\subsubsection{External validity}
This threats refers to the generalizability of the experiments and 
\textit{VIPer}. We mitigated this threat by conducting a large-scale study including five recent and representative vulnerability detectors following token-based and graph-based approaches.
\footnote{We additionally investigated the possibility of using a sixth vulnerability detector, DeepVD\cite{wang2023deepvd}. However, we were unable to reproduce DeepVD's published results and did not receive a response from its authors when we asked for clarification. For this reason, we excluded DeepVD from our study.}
Although, the dataset employed in this study only contains C/C++ code samples, \textit{VIPer}'s primary goal is to ensure accurate representation of vulnerabilities that should be learnt by the detectors, thus, it is language agnostic and dataset independent. Another threat may arise from our reliance on the SARD dataset when devising \textit{SF}s. However, SARD is the largest available dataset that is widely used and contains many real-world security flaws. To further confirm that our results are not inadvertently impacted by SARD, we also examined how the graph-based \textit{SF}s influence ReVeal by utilizing the FFmpeg+Qemu dataset. Our findings indicate that both \textit{SF}s also affect ReVeal's predictions on FFmpeg+Qemu, in a manner consistent with our SARD-based results. This supports the \textit{SF}s' broad applicability.


\subsubsection{Internal validity}
This threat may arise from a weak research protocol or selection bias. We overcome this threat by strictly following the reproducible guidelines provided by the authors of the employed vulnerability detectors. Also, the decision to opt for SARD dataset and specific vulnerability detectors was made based on a comprehensive literature analysis. SARD is relatively the largest dataset available and the chosen detectors are widely employed in prior work.

\subsubsection{Construct validity}
One of the critical design decisions made in this study is to restrict \textit{SF}s for graph-based approaches to two \textit{Node Set} and \textit{Edge Set} in the taxonomy, however, there might be other spurious features that the detectors could be learning from. That said, the taxonomy can be further extended as needed. Another construct validity threat may arise due to the employed evaluation metrics. When reproducing the vulnerability tools, we utilized their adopted evaluation metrics. In contrast, the metric ``satisfaction rate'' is unique to this study which is employed to measure the extent to which the vulnerability detectors deviate from the desired outcomes when FPP and FEP are applied to the code samples.

\subsubsection{Conclusion validity}
This threat concerns with the authenticity and significance of the findings reported in this study. We mitigated this threat by rigorously following authors' guidelines while reproducing the results for their vulnerability detectors. Also, due to absence of replacement for system read APIs, we do not generate FPPs for the \textit{VF} \textit{Read API} and hence did not include in our results. Another threat may be the use of  Joern to extract the CPG from each dataset sample. Joern has been observed to occasionally miss dependencies. However, its developers have actively tried to address this issue in recent updates. In addition to using an updated version, we mitigated this threat by filtering out the invalid CPGs constructed by Joern.

\section{Related work}\label{sec:lit}
Understanding the features relevant to vulnerability prediction (i.e., \textit{VF}s) is critical for improving the DL-based vulnerability detectors' reliability and trustworthiness. Recent studies have, therefore, attempted to explore the \textit{VF}s in DL-based vulnerability detectors. Risse et al.~\cite{risse2023limits} studied the vulnerability detectors’ ability to learn fixes for vulnerable patches. Suneja et al.~\cite{suneja2021probing} probed the signal-awareness of models used for vulnerability detection by reducing the input source code to the minimum tokens required to retain the prediction.
Meanwhile, our approach for devising \textit{VF}s is primarily focused on properties related to code dependencies. It is also the first to evaluate whether vulnerability detection techniques understand these code dependencies.

\looseness=-1
Another way researchers pursue explaining the predictions of DL-based models is by examining the unintended pattern in the dataset that the models might learn erroneously (i.e., \textit{SF}s). On that note, recent studies have explored the influence of variable names~\cite{yang2022natural,rahman2023towards} and method names~\cite{pour2021search,rahman2023towards} in the prediction of the models of code. Risse et al.~\cite{risse2023limits} analyzed the behavior of models of code on logic-preserving transformations.
Our work is the first to categorize \textit{SF}s based on how they impact token-based and graph-based approaches, and to develop \textit{SF}s based on graph properties of code. Additionally, this paper presents the dichotomy between \textit{VF}s and \textit{SF}s through an illustrative taxonomy as well as an evaluation framework that allows for the assessment of both \textit{VF}s and \textit{SF}s within the same platform.

\section{Conclusion and Future Work}
\label{sec:conclusion}
Previous studies have shown that existing vulnerability detectors fail to generalize well to unseen datasets, suggesting they may be learning irrelevant code features. To address this, we introduce \textit{VIPer}, a framework for accurately evaluating vulnerability detectors by providing a comprehensive view of the features that truly contribute to vulnerabilities and how they impact the predictions of vulnerability detectors. \textit{VIPer} employs feature-preserving and feature-eliminating perturbations to assess a detector's performance. Our results reveal that, in the case of vulnerability features, approximately \textasciitilde2\% of feature-preserving perturbations and \textasciitilde84\% of feature-eliminating perturbations have an adverse impact on detector outcomes. In the case of spurious features, we observe that feature-preserving perturbations produced a drop in recall up to $29\%$ for graph-based detectors. To facilitate correct evaluation and improvement of vulnerability detectors, we have made \textit{VIPer} publicly available to the research community. For future work, we plan to discover more vulnerability and spurious features and explore ways to ensure that vulnerability detectors handle them properly.


\section*{Acknowledgment}
This work is supported in part by the National Science Foundation under grant CCF-2106871.

\balance
\bibliographystyle{IEEEtran}
\bibliography{main}


\end{document}